\documentclass[sigconf]{acmart}
\graphicspath{{../figures}} %goes to path: figures/

\AtBeginDocument{%
  \providecommand\BibTeX{{%
    \normalfont B\kern-0.5em{\scshape i\kern-0.25em b}\kern-0.8em\TeX}}}

\setcopyright{acmcopyright}
\copyrightyear{2023}
\acmYear{2023}
\acmDOI{XXXXXXX.XXXXXXX}

\acmConference[HRI '23]{Make sure to enter the correct
  conference title from your rights confirmation emai}{March 13--16,
  2023}{Stockholm, SE}
\acmBooktitle{HRI '23: ACM/IEEE International Conference on Human-Robot Interaction, 
March 13-16, 2023 Stockholm, SE} 
\acmPrice{15.00}
\acmISBN{978-1-4503-XXXX-X/18/06}

\begin{document}

\title{
%How to teach bias in AI and Robotics to Mexican Children?%Sun 15 Jan 07:22:56 GMT 2023
% Teaching bias in AI and Robotics to Mexican Children%Sun 15 Jan 08:25:41 GMT 2023
%Teaching bias in AI and Robotics to Children in a small community in Mexico
%Fri 20 Jan 06:15:26 GMT 2023
Teaching AI and Robotics to Children in a Mexican town
%Mon 23 Jan 23:57:27 GMT 2023
}

\author{Antonio Badillo-Perez}
\affiliation{%
  \institution{air4children}
  \city{Xicohtzinco, Tlaxcala}
  \country{M\'exico}}

\author{Donato Badillo-Perez}
\affiliation{%
  \institution{air4children}
  \city{Xicohtzinco, Tlaxcala}
  \country{M\'exico}}
  
% \author{Antonio Badillo-Perez}
% \affiliation{%

\author{Alex Barco}
\affiliation{%
  \institution{University of Deusto} %Engineering Faculty, 
  % \streetaddress{1 Th{\o}rv{\"a}ld Circle}
  \city{Bilbao Biscay}
  \country{Spain}}

\author{Rocio Montenegro}
\affiliation{%
  \institution{Family Montessori School Society}
  % \streetaddress{8600 Datapoint Drive}
  \city{Vancouver, British Columbia} 
  % \state{Texas}
  \country{Canada}
  % \postcode{78229}
  }

\author{Miguel Xochicale}
\affiliation{%
  \institution{air4children}
  \city{Xicohtzinco, Tlaxcala}
  \country{M\'exico}}
\email{air4children@gmail.com}

\renewcommand{\shortauthors}{Xochicale M. et al.}

\begin{abstract}
In this paper, we present a pilot study aiming to investigate the challenges of teaching AI and Robotics to children in  low- and middle-income countries.
Challenges such as the little to none experts and the limited resources in a Mexican town to teach AI and Robotics were addressed with
the creation of inclusive learning activities with Montessori method and open-source educational robots.
 % teaching bias in AI and
For the pilot study, we invited 14 participants of which 10 were able to attend, 6 male and 4 female of (age in years: mean=8 and std=$\pm$1.61) and four instructors of different teaching experience levels to young audiences.
We reported results of a four-lesson curriculum that is both inclusive and engaging. 
% with Montessori method and low-cost open source educational robots. 
We showed the impact on the increase of general agreement of participants on the understanding of what engineers and scientists do in their jobs, with engineering attitudes surveys and Likert scale charts from the first and the last lesson.
We concluded that this pilot study helped children coming from low- to mid-income families to learn fundamental concepts of AI and Robotics and aware them of the potential of AI and Robotics applications which might rule their adult lives.
Future work might lead (a) to have better understanding on the financial and logistical challenges to organise a workshop with a major number of participants for reliable and representative data and (b) to improve pretest-posttest survey design and its statistical analysis.
The resources to reproduce this work are available at \url{https://github.com/air4children/dei-hri2023}.
\end{abstract}
%of teaching AI and Robotics in a community where little to none experts were able to taught such subjects, and the challenges of creating a curriculum 




\begin{teaserfigure}
    \includegraphics[width=\textwidth]{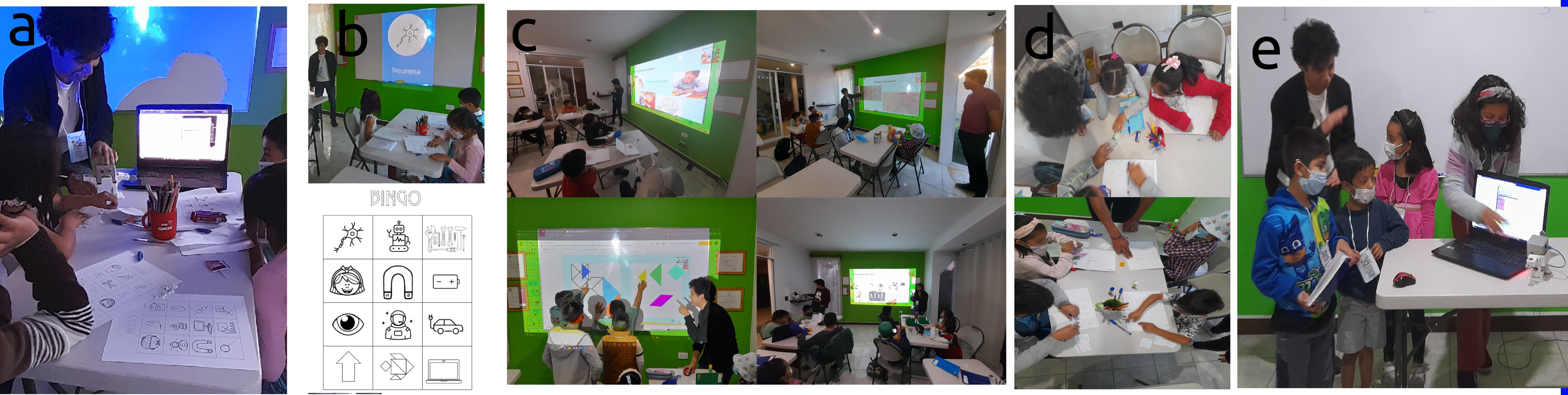} %%OVERLEAF OR ARXIV
    \caption{
    (a) Instructor demonstrating coding activities to children with open-source educational robot,
    (b) instructor playing  bingo game with AI and Robotic-based cards,
    (c) instructors presenting AI and Robotics fundamentals to children,
    (d) children interacting in group activities, and 
    (e) showcase activity where instructor(s) helped a group of four children to present their projects. 
    }
    %\Description{}
    \label{fig:main}
\end{teaserfigure}

%%%%%%%%%%%%%%%%%%%%%%%%%%%%%%%
%% History of the paper
%\received{23rd January 2023}
%\received[revised]{6th Feb 2023}
%\received[accepted]{1st March 2023}

\maketitle

\section{Introduction}
Teaching Artificial Intelligence (AI) and Robotics to young learners has been made good progress in the last decade~\cite{bers2019, druga2019}.
Educational progress has been possible due to the great investment in AI and Robotics in countries such as U.S.A, Germany, Denmark, and Sweden, making those countries and their communities to define core values in AI and Robotics~\cite{druga2019}. 
These raise the question on how to adapt such core values from the most to the least privileged environments~\cite{pratyusha2020}. 
Recently, it has been few progress on using open-source  educational robots and the creation of child-centred programs to teach AI and Robotics in low- and middle-income countries~\cite{abadilloperez2022_DEI_HRI2022, montenegro2021air4children}.
However, there are various challenges on teaching AI and Robotics to young learners.
For instance, the barrier of teaching programming skills to those who are still learning writing and reading skills to which visual and auditory programs using sensor, action and logic blocks helped to address such challenges~\cite{long2020, wyeth2008}.
Other challenge is the little to none experts to teach AI and Robotics and its limited educational material in low- and middle-income countries~\cite{yang2022, abadilloperez2022_DEI_HRI2022}.
% Even though open-source educational robots seems to be affordable, there seems to be
% Current research lines in AI and Robotics claims to be neutral where data pipelines align with principles of findability, accessibility, interoperability, and reusability (FAIR). %TOADD_REF
Hence, the aim of this work is to investigate further challenges of teaching AI and Robotics to young audiences in a Mexican town where little no experts can teach such subjects with limited resources.

%% Here to add a brieft description of the sections of the paper 
In this paper, we present challenges of teaching AI and Robotics to Children in a Mexican town.
In section 3, we present the study design and curriculum design with four lessons.
In section 4, we present results and surveys from a pilot experiment.
We then conclude the paper and add future work in section 5.

%%%%%%%%%%%%%%%%%%%%%%%%%%%%%%%%%%%%%%%%%%%%%%%%%%%%%%%%%%%%%%%%%%%%%
%% VICTOR to review papers and add summary on the challenges on AI literacy
%Touretzky, D., Gardner-McCune, C., Martin, F. and Seehorn, D. (2019) %Envisioning AI for K12: What should every child know about AI? The Thirty-%Third AAAI Conference on Artificial Intelligence (AAAI-19).
%Introducing the Fundamentals of Artificial Intelligence to K-12 Classrooms %According to Educational Neuroscience Principles
% Perhaps cite papers in the 
%https://www.bcu.ac.uk/education-and-social-work/research/news-and-events/cspace-conference-2021/blog/ai-literacy-the-role-primary-education
%
%See https://stefania11.github.io/assets/pdf/FABLEARN_Inclusive_AI_2019.pdf
%\cite{druga2019} 
%
%%%%%%%%%%%%%%%%%%%%%%%%%%%%%%%%%%%%%%%%%%%%%%%%%%%%%%%%%%%%%%%%%%%%%

\section{Challenges of teaching AI and Robotics to Children}
\subsection{Teaching bias in AI}
Smith et al. pointed out the current challenges of when and how to teach computing ethics to students~\cite{smith2022incorporating}.
Learning ethics in introductory courses help students to think ethically about their computation topics~\cite{fiesler2021}.
% TOREVIEW \cite{weerts2022}
Payne et al. have shown great progress on making pilots and beta test teaching ethical AI to young audiences~\cite{payne2020}.
For instance, on summer 2021 a pilot to teach ethical AI to young audiences were organised with 28 kids at the Media Lab with a cost of \$150 for the week, leading to a beta test with 250 students in autumn 2021. 
Such pilots considered question related to the everyday life of children, such as "What's is the best algorithm to make a peanut-butter sandwich?: is it the algorithm that makes the tastiest sandwich? , is it the prettiest sandwich?, is it the quickest and easiest to make? or the easiest to clean up?". 
Hence the challenge is to contextualise such questions in a Mexican town where children may know little to none about AI, Robotics end ethics (bias).
% https://www.wgbh.org/news/science-and-technology/2019/07/31/teaching-kids-the-ethics-of-artificial-intelligence
% https://qz.com/1700325/mit-developed-a-course-to-teach-tweens-about-the-ethics-of-ai

% Immediately following this sentence is the point at which
% Table~\ref{tab:freq} is included in the input file; compare the
% placement of the table here with the table in the printed output of
% this document.
% \begin{table}
%   \caption{Frequency of Special Characters}
%   \label{tab:freq}
%   \begin{tabular}{ccl}
%     \toprule
%     Non-English or Math&Frequency&Comments\\
%     \midrule
%     \O & 1 in 1,000& For Swedish names\\
%     $\pi$ & 1 in 5& Common in math\\
%     \$ & 4 in 5 & Used in business\\
%     $\Psi^2_1$ & 1 in 40,000& Unexplained usage\\
%   \bottomrule
% \end{tabular}
% \end{table}

% Immediately following this sentence is the point at which
% Table~\ref{tab:commands} is included in the input file; again, it is
% instructive to compare the placement of the table here with the table
% in the printed output of this document.

% \begin{table*}
%   \caption{Some Typical Commands}
%   \label{tab:commands}
%   \begin{tabular}{ccl}
%     \toprule
%     Command &A Number & Comments\\
%     \midrule
%     \texttt{{\char'134}author} & 100& Author \\
%     \texttt{{\char'134}table}& 300 & For tables\\
%     \texttt{{\char'134}table*}& 400& For wider tables\\
%     \bottomrule
%   \end{tabular}
% \end{table*}

%%%%%%%%%%%%%%%%%%%%%%%%%%%%%%%%%%%%%%%%%%%%%%%%%%%%%%%%%%%%%%%%%%%
%%CHIO please help to review, to change or to add to the following section
\subsection{Inclusive learning of AI and Robotics with Montessori method}
Montessori method is focused in child-centred development where the role of Montessori teachers is to help children individually or in small groups engaging in self-directed activities \cite{Aljabreen2020}.
The adoption of technological topics (AI and Robotics) into Montessori methods has been experimented and studied in the last 10 years \cite{elkin2014}. 
However, there are other skills that are also required to create inclusive learning of new technologies. For example, "collaborative skills" and "understating concepts" are two main factors to design inclusive AI literary for children in low, medium and high socioeconomic backgrounds \cite{druga2019}. 
Also, there is evidence on the impact of collaboration and engagement activities on how coding games and robots to enhance computational thinking \cite{sharma2019}.
Hence, the challenge is how to adapt Montessori method to not only create conditions to develop physical and social requirements but collaborative and literacy skills through hands-on activities. 

\begin{figure}[t]
  \centering
    \includegraphics[width=\linewidth]{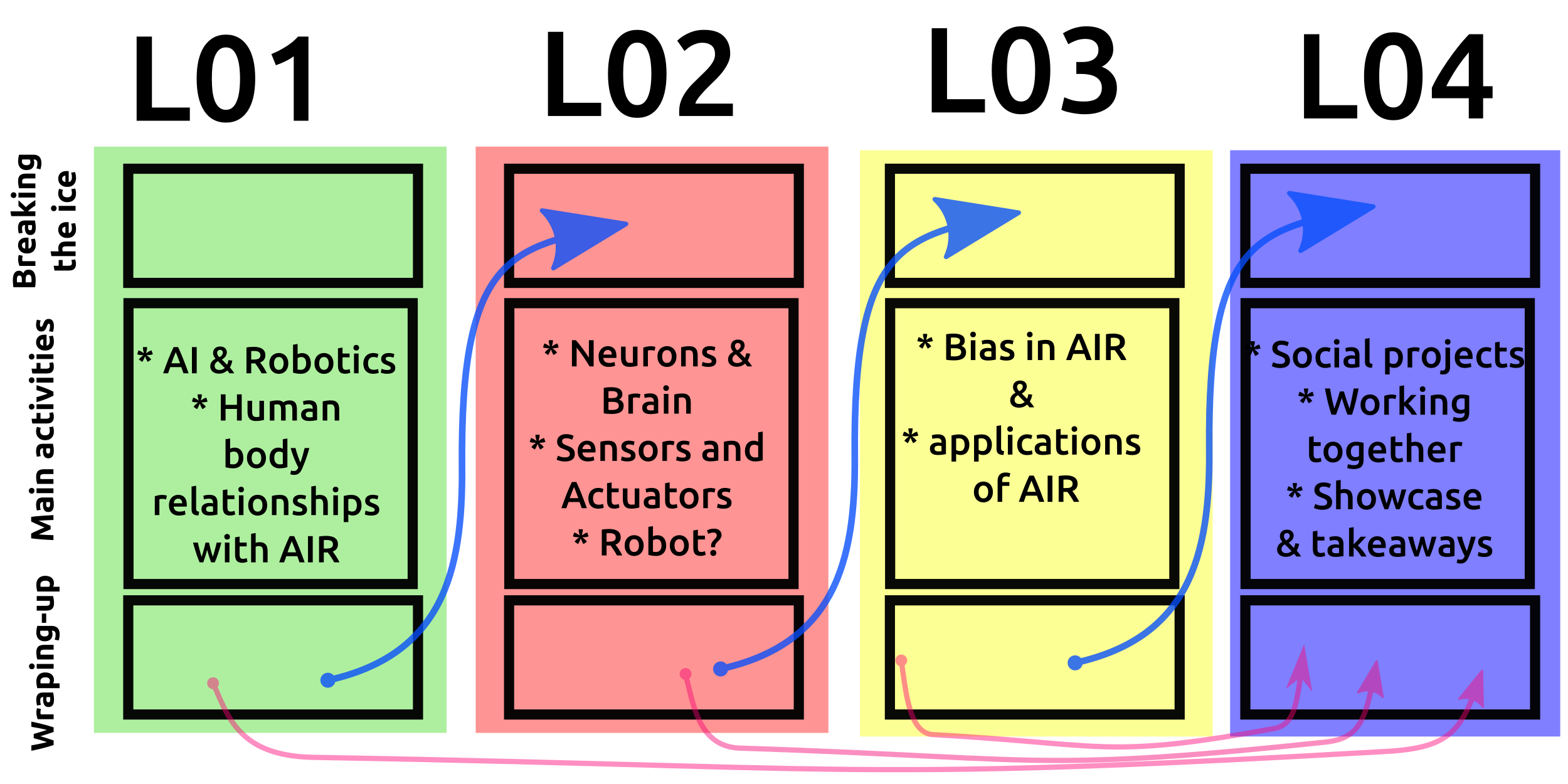}  %%OVERLEAF and ARXIV
    \caption{
    Curriculum design with four lessons of 1 hour and a half (L01, L02, L03 and L04).
    The arrows illustrate the connection between the the last and the first part of each lesson as way to summarise the progress of each lesson (blue arrows) and connections with their full curriculum (red arrows).
    }
    \label{fig:curriculum}
\end{figure}
\section{Pilot study: study design and curriculum design}
%%%%%%%%%%%%%%%%%%%%%%%%%%%%%%%%%%%%%%%%%%%%%
% Brieft introduction to this section [MX/TB]
\subsection{Participants}
In this pilot study, we invited 14 participants between the age of 6 to 9 years old. 
During the pilot workshop, only 10 participants 6 male and 4 female of (age in years: mean=8 and std=$\pm$1.61) were able to join the workshops as remained ones were unable to attend due to personal circumstances. 
We also invited four instructors with various experiences levels in teaching subjects to young audiences (two between 5 to 10 years and two with less than 2 years) and one person to lead the logistics of the event.

\subsection{Lessons}
Aiming to create inclusive teaching practices for students, we considered a study design based on procedures, instruments,  data collection and analysis~\cite{du2022}.
Considering the limited human and material resources, we decided to teach four lessons instead of the original ten lessons for the pilot study of this curriculum.

%%%%%%%%%%%%%%%%%%%%%%%%%%%%%%%%%%%%%%%%%%%%%
%% TON~O and DONATO will draft details of the lessons in four paragraphs (or one large paragraph) 
Figure~\ref{fig:curriculum} illustrates the curriculum design for the pilot experiment. 
During the first lesson, we introduced basics concept of human senses and how AI relates to identifying cats and dogs. We were able to use more dynamic activities with protoboards and simple electric circuits so that children can easily learn fundamentals on how a robot works (Figure~\ref{fig:main}d). 
For the second lesson, we presented examples on how AI learn to recognize animals or humans with a tangram activity and gave them a mini introduction to the code that they will use for the robot.
In the third lesson, we adopted a bingo game with key terms from the previous lessons (Figure~\ref{fig:main}b). We 
introduced what is algorithm with an activity of cooking quesadillas where students learnt that everyone can have a different answer given the same instruction.
We also organised activities on action/reaction with different sensors, how sensors work, how to program sensors and examples with the protoboards and circuits.
During the fourth lesson, we closed the course, starting with an activity in the garden where students program their teammates and practice to give instructions to a robot in order to get the wanted result. Then we show them some real life applications of AI and robotics and we grouped participants in three teams so they could program their robots and design their logo (Figure~\ref{fig:main}e). 

\subsection{Surveys}
Considering "Engineering Attitudes Survey" \cite{cunningham2010impact}, we surveyed all participants with same survey, one for the first lesson and one for the last day of the lessons. 
Considering the reduction of time, fatigue and financial compensation, we consider a three-scale Likert chart \cite{jebb2021}, where each statement in our survey has three numbers where 1 is totally disagree; 2: not sure, and 3 totally agree .
See original statements in Table~\ref{tab:questions} and Spanish translation of the statements in Fig~\ref{fig:survey}.

To analyse the results of the survey, we consider the percentages of the responses of each statement from the first and the last lesson and illustrate the results using percentage response of Likert scale charts.
Considering the small sample of participants and the no assumptions to data distributions, we choose the nonparametric Wilcoxon signed-rank test for statistical analysis for answers of pretest and posttest survey \cite{scipy2001}.

\section{Results}
% \subsection{Pilot study}
Four lessons of the designed curriculum were organised on Mondays and Wednesdays in a Mexican town during two weeks of November 2022. 
Instructors helped children in the pilot workshop by creating groups of three to four children.
In the final lesson groups presented their project with their own robot and children with the help of the instructor explained the application of the robot (Figure~\ref{fig:main}).

\begin{figure}[t]
  \centering
    \includegraphics[width=\linewidth]{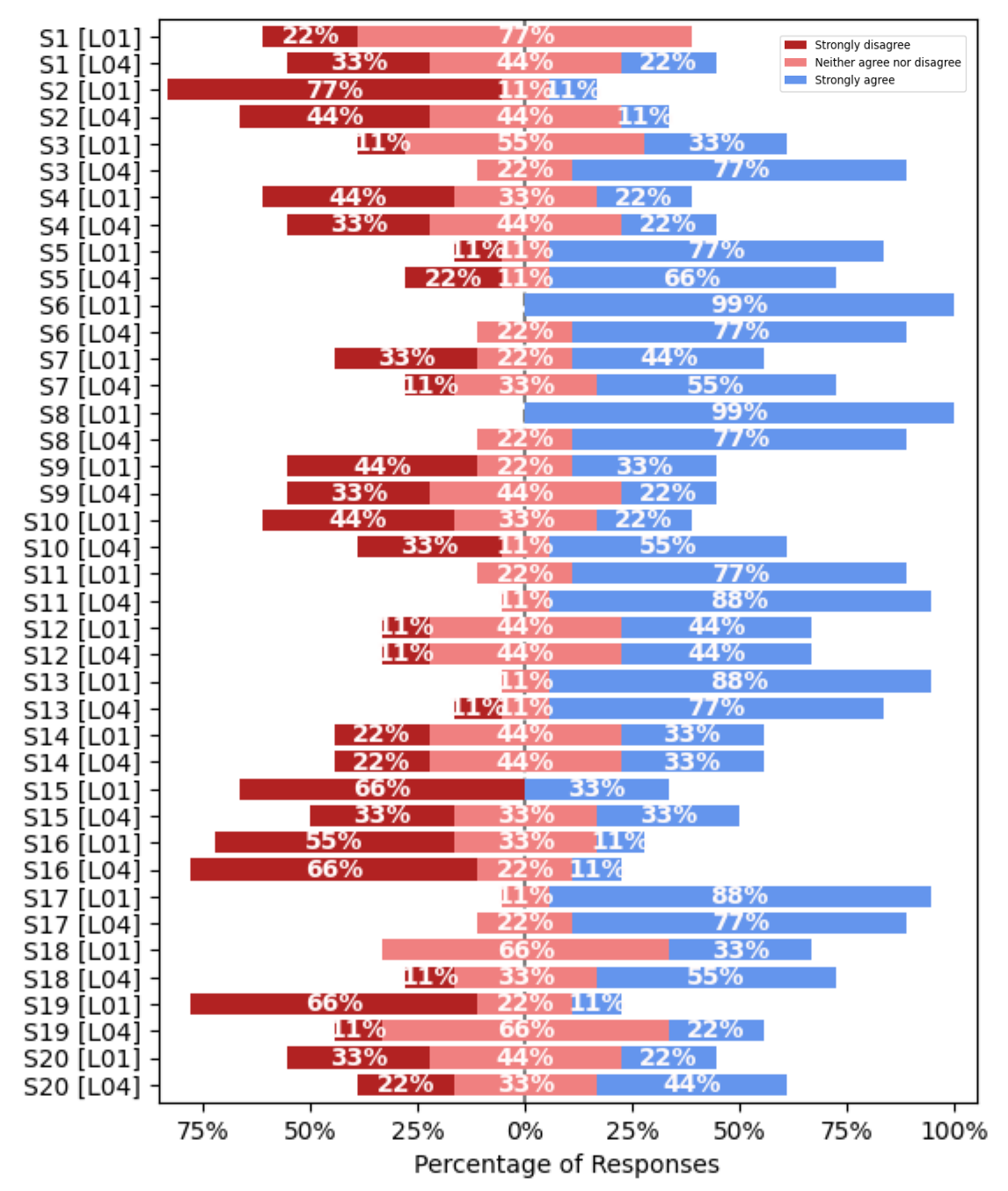}  %%OVERLEAF and ARXIV
    \caption{
    Percentages of responses of 9 participants from the survey with 20 statements (S1 to S20) of the first lesson (L01) and the fourth lesson (L04).
    %"S17: Scientist help make people’s lives better." resulted in the highest average value whereas 
    %"S16: Engineers cause problems in the world." is the lowest average value.
    % "S19: I think I know what scientists do for their jobs." and "S20: I think I know what engineers do for their jobs." shows an increase on agreement. 
    See Appendices for further references on the statements of the survey and reproducible jupyter notebooks.
    }
    \label{fig:results}
\end{figure}

We only considered 9 participants for the survey as 5 participants were neither in one of the days of the surveys or there were not able to be in person in such days. 
Figure~\ref{fig:results} illustrates percentages or responses of 20 statements from 9 participants for 1 which is totally disagree; 2: not sure, and 3 totally agree.
It can be noted that "S3: I would like a job where I could invent things.", "S11: I would like a job that lets me figure it out how things work"
and ", and "S18: Engineers help make people’s lives better." showed an increase of agreement. 
Whereas "S6: I would like to build and test machines that could help people to walk." and  "S8: I would enjoy a job helping to protect the environment." created less disagreement.
"S17: Scientist help make people’s lives better." resulted in the highest average value whereas "S16: Engineers cause problems in the world." is the lowest average value.
"S19: I think I know what scientists do for their jobs." and "S20: I think I know what engineers do for their jobs." showed both an increase on agreement between participants and decrease in disagreement.

A Wilcoxon T test was used to analyze the results of the survey before and after the survey to see if the engineering attitudes had a significant effect on pre and post survey of the workshop.
The average survey before the test was lower ($\mu$ = 2.194500 $\pm \sigma$ 0.558367 ) compared to the posttest results ($\mu$ = 2.239500 $\pm \sigma$= 0.396796).
There was no statistically significant in the increase of attitudes towards engineering (t=53.5, p= 0.45).
See Appendix section for reproducible Jupyter notebooks of statistical analysis and plots.

% S11: I would like a job that lets me figure it out how things work. 
% S13: I like knowing how things work. 
% Wheres 

% \subsection{Surveys}
% TODO

\section{Conclusions and future work}
%%%%%%%%%%%%%%%%%%%%%%%%
%%% CONCLUSIONS
%%% Please add any conclusions that you saw during the pilot
In this paper, we investigated the challenges of teaching AI and Robotics to young audiences with a pilot study in a Mexican town.
We designed ten lessons but only were able to pilot four lessons due the limited material and human resources. 
We also surveyed participants using Engineering Attitudes Survey from \cite{cunningham2010impact}, and showed the impact on the increase of general agreement of understanding what engineers and scientist do in their jobs using Likert scale charts.
However, statistical analysis suggested no statistically significant in the increase of attitudes towards
engineering  which we think is related to our small sample. 
Also, during the survey we noted that children were not able to read long sentences and were confused about agreement and disagreement terminology in the Likert scale.

We conclude that children and parents had the willingness to join this pilot study by learning new experiences on collaborative and fun activities
based on Montessori education and with low-cost educational robots. 
We also conclude that to be inclusive we would like to include anyone from a potential generation that might be ruled by AI. 
Hence, thanks to these activities we allowed these children coming from low- to mid-income families to learn and be aware of the potential of AI which is crucial in this technological world.

However, the limited and self-funded budget for this project restricted us to prepare and organise a major number of lessons and reach more participants which hopefully will be addressed in future work. 
We will also improve our pretest-posttest survey design with a reliable and representative data and its statistical analysis.

%%%%%%%%%%%%%%%%%%%%%%%%
%%%FUTURE WORK
%%% please add your initials and paragraphs on what you might consider be a good line of research for future work.
%% MX: How to incorporate other examples of bias in AI/Robotics in the context of the Mexican society?

\begin{acks}
% To everyone involved in this collaborative work.
To Andrea Aguilar-Galindo, Gustavo Barbosa, Donato Badillo-Perez and Antonio Badillo-Perez for volunteering as instructors in the workshops.
To Martha P\'erez and Donato Badillo for their support in organising the pilot of the workshops in their teaching installations. 
To Rocio Montenegro for her contributions with the design of activities based on Montessori education.
To Alex Barco for his feedback to improve the scientific impact of our collaborative work.
To Mercedes P\'erez Mendonza for her feedback on interactive activities.
To Victor Alonso for his feedback on teaching AI to children. 
To Leticia V\'azquez for her support with the translation of surveys and feedback to improve the workshops.
To El\'ias M\'endez Zapata and Abdel Rodriguez Cuapio from Tlaxcala Polytechnic University for their support with the invitation of students and feedback on the hardware design of the robot.
To Xiaoxue Du from MIT Media Labs for her suggestions on the study design. 
To Angel Mandujano, Elva Corona and others who have supported AIR4children project to keep it moving forward. 
\end{acks}

\bibliographystyle{ACM-Reference-Format}
%\bibliography{../references/references} %%GITHUB
%\bibliography{references} %%OVERLEAF
%\bibliography{../../references/references} %%ARXIV

 %% uncomment for arxiv version

%%%
%%% If your work has an appendix, this is the place to put it.

\newpage

\section*{Appendices}
\appendix

\section*{Surveys}
\begin{figure} %[h]
  \centering
    \includegraphics[width=\linewidth]{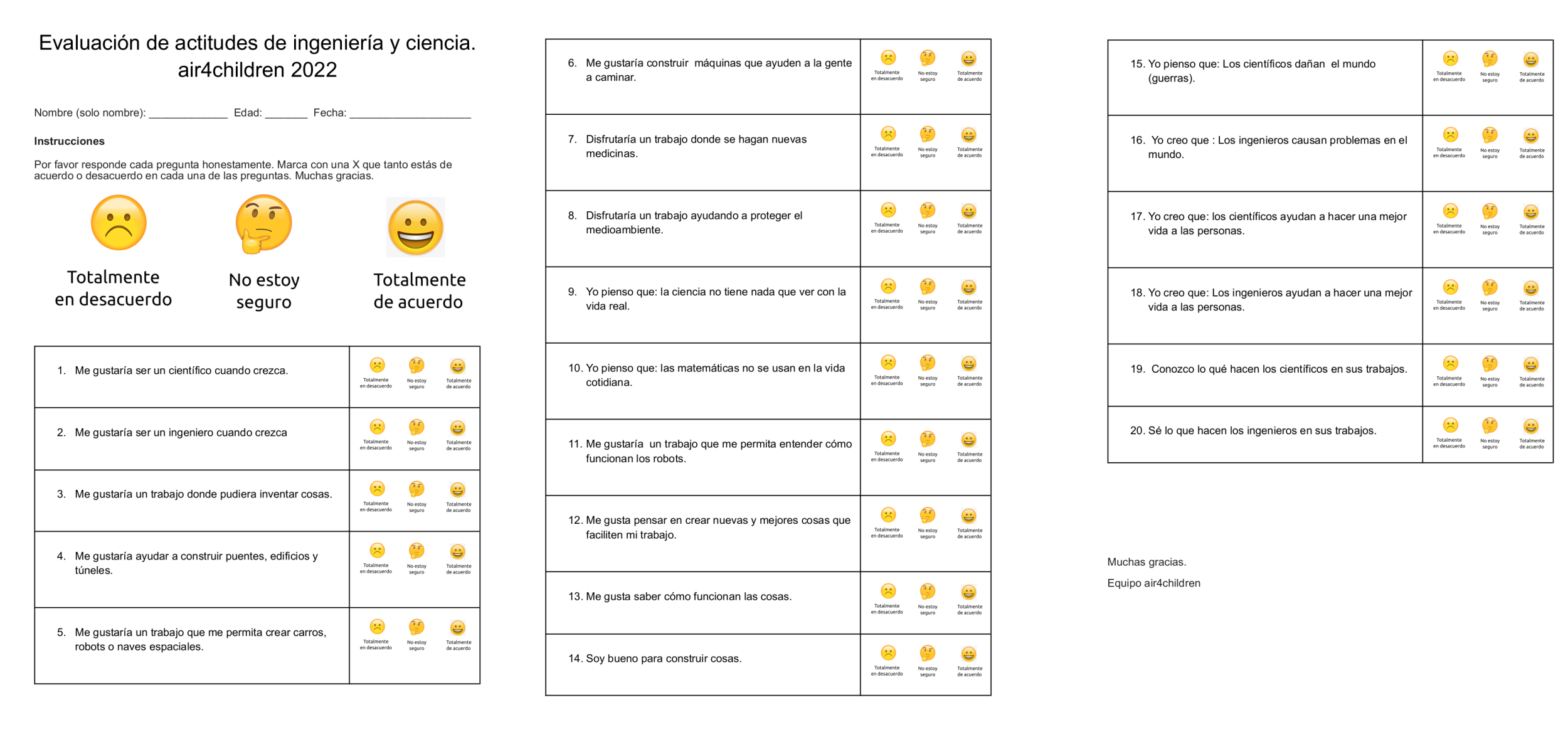}  %%OVERLEAF and ARXIV 
    \caption{
    Survey in Spanish with 20 statements.
	Statements were translated from English to Spanish from the "Engineering Attitudes Survey" \cite{cunningham2010impact}.
    }
    \label{fig:survey}
\end{figure}

Figure~\ref{fig:survey} shows survey layout document and table~\ref{tab:questions} presents 20 statements of the survey.

\begin{table*}[h]
  \caption{Statements in the survey from "Engineering Attitudes Survey" \cite{cunningham2010impact}.
}
  \label{tab:questions}
  \begin{tabular}{ll}
    \toprule
    Number & Statements \\
    \midrule
    S1 & I would enjoy being a scientist when I grow up \\
    S2 & I would enjoy being an engineer when I grow up.  \\
    S3 & I would like a job where I could invent things. \\
    S4 & I would like to help to plan bridges, skyscrapers, and tunnels. \\
    S5 & I would like a job that lets me design a cars. \\   
    S6 & I would like to build and test machines that could help people to walk. \\
    S7 & I would enjoy a job helping to make new medicines. \\
    S8 & I would enjoy a job helping to protect the environment. \\
    S9 & Science has nothing to do with real life. \\
    S10 & Math has nothing to do with real life. \\
    S11 & I would like a job that lets me figure it out how things work. \\
    S12 & I like thinking of new and better ways of doing things. \\
    S13 & I like knowing how things work. \\
    S14 & I am good at putting things together. \\
    S15 & Scientist cause problems in the world. \\
    S16 & Engineers cause problems in the world. \\
    S17 & Scientist help make people’s lives better. \\
    S18 & Engineers help make people’s lives better. \\
    S19 & I think I know what scientists do for their jobs. \\
    S20 & I think I know what engineers do for their jobs. \\
    \bottomrule
  \end{tabular}
\end{table*}

\section*{Jupyter notebooks}
\begin{figure}[h]
  \centering
    \includegraphics[width=\linewidth]{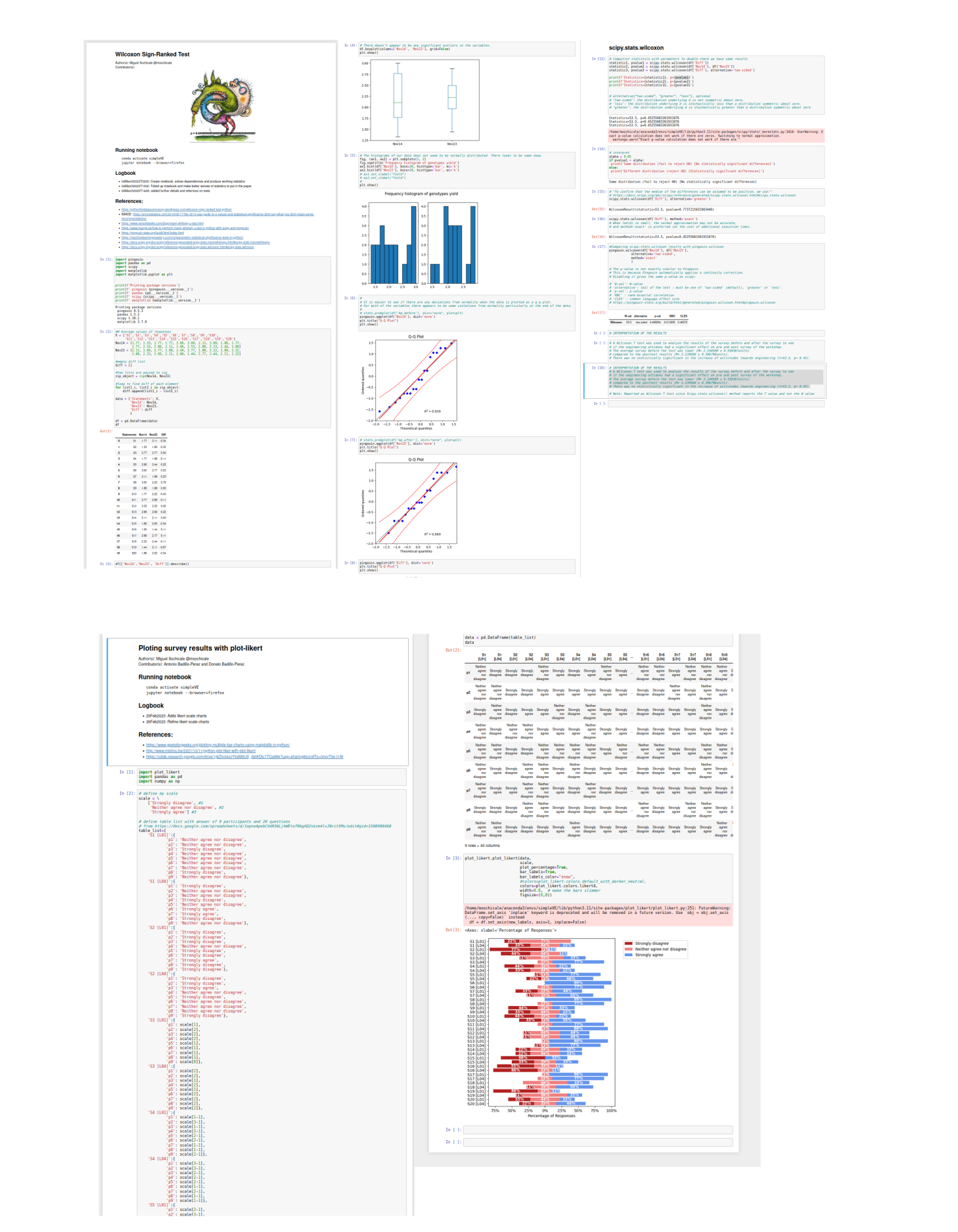}  %%OVERLEAF and ARXIV 
    \caption{
    Jupyter notebooks are available at \url{https://github.com/air4children/dei-hri2023/}
    }
    \label{fig:notebooks}
\end{figure}

Figure~\ref{fig:notebooks} shows Jupyter notebooks which are available at \\ \url{https://github.com/air4children/dei-hri2023/}


%Etiam commodo feugiat nisl pulvinar pellentesque. Etiam auctor sodales
%ligula, non varius nibh pulvinar semper. Suspendisse nec lectus non
%ipsum convallis congue hendrerit vitae sapien. Donec at laoreet
%eros. Vivamus non purus placerat, scelerisque diam eu, cursus
%ante. Etiam aliquam tortor auctor efficitur mattis.
%

\end{document}